\documentclass[12pt,preprint]{aastex}

\input psfig.sty

\slugcomment{ApJ, in preparation}

\shorttitle{}
\shortauthors{Rice \& Armitage}

\begin{document}

\title{Quantifying orbital migration from exoplanet  
statistics \\ and host metallicities}

\author{W.K.M. Rice\altaffilmark{1} and Philip J. Armitage\altaffilmark{2,3}}
\altaffiltext{1}{Institute of Geophysics and Planetary Physics and Department 
of Earth Sciences, University of California, Riverside, CA 92521;
ken.rice@ucr.edu}
\altaffiltext{2}{JILA, Campus Box 440, University of Colorado, Boulder CO 80309; 
pja@jilau1.colorado.edu}
\altaffiltext{3}{Department of Astrophysical and Planetary Sciences, University of Colorado, Boulder CO 80309}

\begin{abstract}
We investigate how the statistical distribution of extrasolar planets 
may be combined with knowledge of the host stars' metallicity to yield 
constraints on the migration histories of gas giant planets. At any radius, 
planets that barely manage to form around the lowest metallicity stars accrete 
their envelopes just as the gas disk is being dissipated, so the lower 
envelope of planets in a plot of metallicity vs semi-major axis defines 
a sample of non-migratory planets that will have suffered less than 
average migration subsequent to gap opening. Under the assumption 
that metallicity largely controls the initial surface density of 
planetesimals, we use simplified core accretion models to calculate 
how the minimum metallicity needed for planet formation varies as a 
function of semi-major axis. Models that do not include core migration 
prior to gap opening (Type I migration) predict that the critical metallicity 
is largely flat between the snow line and $a \approx 6$~AU, with a weak 
dependence on the initial surface density profile of planetesimals. When 
slow Type~I migration is included, the critical metallicity is found 
to increase steadily from 1-10~AU. Large planet samples, that include 
planets at modestly greater orbital radii than present surveys, therefore 
have the potential to quantify the extent of migration in both Type~I and 
Type~II regimes.
\end{abstract}


\section{Introduction}
The discovery of 51 Pegasi \citep{mayor95} provided evidence for the 
potential importance of orbital migration \citep{goldreich80,lin86} in 
determining the structure of extrasolar planetary systems. For 51~Peg, 
and for other members of the class of hot Jupiters, in situ formation 
scenarios face obvious difficulties due to the predicted high temperature 
($T > 10^3 \ {\rm K}$) of the protoplanetary disk \citep{bell97} at 
the radii -- within 0.1~AU of the star -- where the planets now orbit. 
This conclusion is largely borne out by detailed models of giant planet 
formation, which confirm that planet formation via core accretion is 
unlikely at such small distances from the star \citep{bodenheimer00}. 
For the much larger population of extrasolar giant planets which orbit 
at $a > 1$AU, however, the situation is less clear. These planets, 
which are still orbiting their stars at much smaller radii than 
Jupiter's 5.2~AU, {\em could} potentially form in situ via core accretion 
\citep{bodenheimer00}, especially when recent downward revisions to 
both the radial location \citep{sasselov00} and importance \citep{lodders03} 
of the snow line are considered. Alternatively, the typical formation 
radius for giant planets around roughly Solar mass stars could fall 
beyond the radius of any observed extrasolar planet, implying a 
dominant role for migration in presently known systems.

Observationally, the statistical distributions of basic extrasolar planet 
properties (mass $M_p \sin (i)$, semi-major axis $a$, and eccentricity $e$) 
provide only conflicting clues as to the importance of migration. The low 
masses of some close-in planets point to relatively short residence 
times within the gaseous protoplanetary disk, since numerical simulations 
show that Saturn-mass planets accrete gas across gaps and grow to 
larger masses rather promptly \citep{lubow99,bate03}. If mass growth 
is ignored, however, the increasing fraction of known extrasolar planets 
with orbital radius is consistent with numerical models that assume that 
these planets all formed further out, and migrated inward due to 
planet-disk interactions \citep{armitage02}. Theoretically, it is well 
known that the time scale for core accretion to form planets at large radii 
$a \sim 20$AU is long \citep{pollack96} -- at least in the simplest 
versions of the theory -- but where in the inner disk planet formation 
is most favored is uncertain. Theoretical work has demonstrated that the 
time scale and outcome of core accretion can be modified by Type~I 
migration \citep{ward97,tanaka02} of giant planet cores \citep{hourigan84,alibert04}, 
by gravitational interaction of cores with turbulent fluctuations in the 
disk \citep{rice03}, or by competition for planetesimals between several 
growing cores \citep{hubickyj04}. That this is by no means an exhaustive 
list of the possibilities illustrates that additional observational 
clues as to where and when giant planets form would be valuable.
 
In this paper, we investigate how models of giant planet formation via 
core accretion could be constrained by adding observational knowledge of the 
metallicity of the host star. The fraction of roughly Solar-type stars 
that host known extrasolar planetary systems rises rapidly with the 
stellar metallicity \citep{gonzalez98,santos01,santos04,fischer04,fischer05}, 
with several lines of evidence pointing to the high [Fe/H] being the 
cause, rather than the consequence, of planet formation \citep{kornet05}. The strikingly 
strong scaling of planet frequency with metallicity -- which rises from 
almost zero frequency at ${\rm [Fe/H]} < -0.5$ to 
$\approx$20\% at ${\rm [Fe/H]} \approx +0.5$ -- suggests that metallicity 
is a more important parameter in determining the probability of giant 
planet formation than intrinsic dispersion in either the gas disk mass 
or disk lifetime. If true, then the epoch of giant planet formation at 
a given radius in the disk (defined as the moment when the core accretes 
a significant gaseous envelope via runaway accretion) should correlate 
with the metallicity. High metallicity implies a larger surface density 
of planetesimals, and much shorter planet formation time scales \citep{pollack96}. 
In particular, as illustrated in Figure 1, at any 
radius there should be a mimumum threshold or critical metallicity, 
below which core accretion fails to reach runaway within the lifetime 
of the gas disk. These `failed' gas giants potentially constitute 
a large population of $\sim 10 M_\oplus$ planets much closer in than 
Uranus and Neptune in the Solar System \citep{ida04}. More importantly 
for our purposes, planets just {\em above} the critical metallicity 
form just as the gas is being dissipated. There is therefore little or 
no opportunity for these planets to migrate via Type~II gravitational 
interactions with the protoplanetary disk. Moreover, if the critical 
metallicity increases monotonically with increasing semi-major axis 
(and migration is inward, as is very probable at small radii), then 
there is no way to populate the region of parameter space below the 
threshold curve. Observational definition of the lowest metallicity 
stars that host planets at different radii can therefore probe the 
relative radial efficiency of the planet formation process, 
independent of the separate uncertainties that attend the migration 
process.

The outline of this paper is as follows. In \S2, we describe 
simplified models for gas giant planet formation that allow us 
to calculate the time scale for a single core to reach runaway 
gas accretion in the protoplanetary disk. In \S3, we compute 
the threshold metallicity as a function of orbital radius, and 
discuss how it depends upon the surface density profile of 
planetesimals and on the extent of Type~I migration of giant 
planet cores prior to the accretion of the envelope. The 
most significant differences between the models occur at 
relatively large radii which, although poorly sampled today, 
should be accessible to future astrometric and direct 
imaging surveys. In \S4, we analyze existing statistics of 
extrasolar planetary systems within the context of our 
theoretical model. Although useful constraints on planet 
formation are not possible with existing data, we demonstrate 
that one of the basic theoretical premises -- that planets 
observed to lie near the threshold metallicity curve should 
have formed on average at later epochs -- is consistent with  
the observed distribution of planetary masses around stars 
with different [Fe/H].

\section{Giant planet formation model}
We assume that the massive planets observed in extrasolar planetary 
systems formed via core accretion, and that the dominant factor 
controlling whether planets form at a given radius prior to the 
dispersal of the gas disk is the metallicity of the gas that 
formed the star and disk. Observations suggest that in most 
cases the current photospheric stellar metallicity reflects the 
primordial value \citep{santos04b}, and hence we assume that the 
ratio of the surface densities of the planetesimals to that of 
the gas is linearly related to the stellar metallicity. Our 
goal is to compute as a function of radius the minimum metallicity 
that allows massive planets -- defined as those with 
$M_{pl} > 100 M_\oplus$ -- to form within the lifetime of the gas
disk. We describe here a simplified model for core accretion, similar 
to that developed by \citet{ida04}, that accomplishes this objective.

\subsection{The gas disk}
Within core accretion models of giant planet 
formation \citep{bodenheimer86, pollack96}, the time 
dependent evolution of the gaseous component of the protoplanetary disk 
primarily affects the final masses of planets and the rate of Type~II 
migration subsequent to gap formation. The {\em threshold} for planet 
formation, i.e. whether a gas giant can form at all in a given disk, 
does depend upon the properties of the gas at late epochs (via the 
pressure and temperature of the disk, which provide boundary conditions 
for the planetary envelope), but this dependence is weak compared to 
the effect of changes in the planetesimal surface density. A minimal 
description of the gaseous disk therefore requires only specification 
of the surface density profile and lifetime. We adopt for the gaseous 
surface density a power-law over the range of radii ($1 \ AU < a < 10 \ AU$) 
of interest,
\begin{equation}
 \Sigma_g = 7.2 \times 10^3 \left( {a \over {1 \ {\rm AU}}} \right)^{-\beta} 
 \ {\rm g \ cm^{-2}},
\end{equation} 
where $a$ is the orbital radius and $\beta$ is a parameter that specifies 
the mass distribution in the 
disk (note that we fix the gas surface density at a radius of 1~AU). We 
consider values of $\beta$ in the range $1 \leq \beta \leq 2$, which 
includes the most commonly considered possibilities \citep{weidenschilling77,
bell97,kuchner04}. This surface density profile is strictly the 
profile at the time when planetesimals form from the dust within the gas -- since
this is likely to occur at an early time it will differ little from the 
initial profile. We ignore star-to-star variations in the initial disk 
mass \citep{armitage03}. Provided that these are of less importance 
than (and uncorrelated with) variations in gas metallicity they should 
only introduce scatter in the minimum metallicity required for planet 
formation.

For the lifetime of the gas disk, we take $\tau = 5 \times 10^6 \ {\rm yr}$. 
This value is similar to observational estimates \citep{strom89,haisch01}. 
We assume that once $t > \tau$, the gas disk is promptly dispersed at 
all radii and the potential for further gas giant formation is 
quenched. Observations support the idea that dispersal of the gas 
occurs rapidly \citep{wolk96}.

\subsection{The planetesimal disk}
We assume that the surface density of planetesimals, $\Sigma_d$, has a power-law 
distribution such that
\begin{equation}
\Sigma_d = 10 f_{dust} \eta_{ice}  
\left(\frac{a}{1 \mathrm{AU}}\right)^{-\beta} \
\mathrm{g \ cm^{-2}},
\label{sigma_dust}
\end{equation}
where $\beta$ has the {\em same} value for the planetesimals as for the 
gas\footnote{We note that although this is the simplest assumption, it 
might not be correct, in particular if there is significant radial 
migration of solids prior to planetesimal formation. Recent models 
of planetesimal formation via gravitational instability 
\citep{goldreich73,youdin02,youdin04} require such migration.}. 
We consider $\beta$ values of 
$1$, $1.5$, and $2$. $f_{dust}$ is a parameter that we vary to change 
the normalization of the planetesimal surface density (and stellar 
metallicity), and $\eta_{ice}$ is a step-function that represents the ice 
condensation/sublimation across the snow line which occurs, around a star
of mass $M_*$, at $a_{ice} = 2.7 (M_*/M_\odot)^2$ AU. Since we will
be considering the formation of planets around solar-like
stars, we assume $M_* = M_\odot$ and therefore $a_{ice} = 2.7$ AU. 
Following \citet{ida04} we take $\eta_{ice} = 1$ when $a < a_{ice}$, 
and $\eta_{ice} = 4.2$ when $a > a_{ice}$. 
 
To convert between our parameter $f_{dust}$ and the stellar [Fe/H], we 
note that if $f_{dust} = 3$, then the metallicity is solar. If $f_{dust}$ 
is greater/less than $3$ the metallicity is greater/less than solar. The 
opacity of the gas is also important, since there is a weak dependence
of the critical core mass, and a stronger dependence of the envelope's 
Kelvin-Helmholtz time, on $\kappa$. We assume that,
\begin{equation}
 \kappa = {f_{dust} \over 3} \ {\rm cm^2 g^{-1}}
\end{equation}
so that for solar metallicity the opacity is $1 \ {\rm cm^2 g^{-1}}$. 

\subsection{Core accretion model}
To calculate the growth of a giant planet core within the above disk 
model, and the subsequent accretion of a gaseous envelope, we employ 
a modified version of the scheme developed by \citet{ida04}. For 
each run, we start our calculation with a solid core of mass
$M_{core} = 10^{-3} M_\oplus$, density of $3.2 \mathrm{\ g \ cm^{-3}}$, 
and radius $a$. The core grows at the rate $\dot{M}_{core} = \pi R_c^2 \Sigma_d \Omega F_g$, 
where $R_c$ is the radius of the core, $\Omega$ is the angular frequency,
and $F_g$ is the gravitational enhancement factor \citep{greenzweig92}. The
core accretes planetesimals from a annular region around it known as the 
`feeding zone' that has a full width of $\Delta a_c = 8 r_H$, where
\begin{equation}
r_H =\left(\frac{M_{pl}}{3 M_*}\right)^{1/3} a
\label{rH}
\end{equation}
is the Hill radius for a planet with mass $M_{pl}$, around a star with mass
$M_*$.

As the core grows we adjust, accordingly, the planetesimal surface density
in the feeding zone. In this way the core self-consistently stops growing
once it reaches its isolation mass. In our initial calculations we assume 
that the core's orbital radius remains fixed, though subsequently we 
relax this restriction and allow for orbital migration. 

When the core's mass exceeds a critical value, $M_{crit}$, it is no longer able to
support a gaseous envelope in quasi-hydrostatic equilibrium, and
runaway gas accretion occurs \citep{mizuno80, bodenheimer86,papaloizou99}. This 
critical core mass depends on the planetesimal accretion rate, $\dot{M}_{core}$, 
and on the opacity, $\kappa$, associated with the disk gas. We adopt a 
representative estimate for the critical core mass calculated by 
\citet{ikoma00}
\begin{equation}
M_{crit} = 10 \left(\frac{\dot{M}_{core}}{10^{-6} M_\oplus \mathrm{yr}^{-1}}\right)^
{0.25}\left(\frac{\kappa}{1 \mathrm{\ cm^2 \ g^{-1}}}\right)^{0.25} M_\oplus.
\label{Mcrit}
\end{equation}
Once the core exceeds the critical core mass, the gaseous
envelope contracts on a Kelvin-Helmholtz timescale, $\tau_{KH}$. 
\citet{ikoma00} show that the Kelvin-Helmholtz timescale can be written as
\begin{equation}
\tau_{KH}=10^b\left(\frac{M_{pl}}{M_\oplus}\right)^{-c}\left(\frac{\kappa}{1 \mathrm{\ g \ cm^{-2}}}\right) \mathrm{yr}
\label{tKH}
\end{equation}
where the exact values of $b$ and $c$ depend
on the choice of opacity table. \citet{ikoma00} found that 
$b \simeq 8$ and $c \simeq 2.5$, while \citet{bryden00} obtained a 
fit to the results of \citet{pollack96} with $b \simeq 10$ and $c \simeq 3$. 
We follow \citet{ida04} and use $b = 9$ and $c = 3$. Once the critical core 
mass has been exceeded, we allow gas accretion, at a rate, 
\begin{equation}
\frac{dM_{pl}}{dt} = \frac{M_{pl}}{\tau_{KH}},
\label{dMpl}
\end{equation}
where $M_{pl}$ includes the mass of both the solid core and the gaseous envelope.
It should be noted that in this model the core can continue to grow while gas is being added.

Equations (\ref{tKH}) and (\ref{dMpl}) define a model for growth of a giant planet's 
envelope that is entirely demand-driven -- there is no dependence whatsoever on the 
properties of the gas disk. This is reasonable for low mass planets, but will fail 
at higher masses when either the supply of gas becomes limited or the planet opens 
a gap. Since we are interested in planets that are forming just
as the disk gas is dissipating, we assume that there is sufficient supply 
for these planets to reach masses of $\sim 100 M_\oplus$. If runaway growth does occur in 
our model, we stop the calculations when $M_{pl} > 100 M_\oplus$. This mass 
threshold (which corresponds to 0.3 Jupiter masses) defines `success' in 
forming a giant planet.    

The procedure for modeling planet growth is then as follows. At $t=0$ we start 
with a $10^{-3} M_\oplus$ core located at a radius $a$. We calculate $\dot{M}_{core}$ 
which we use to determine the core mass at the next timestep. Simultaneously we determine $M_{crit}$. 
If $M_{core} > M_{crit}$ then gas is accreted with the contraction timescale given by $\tau_{KH}$. 
We stop the calculation once $t > 5 \times 10^6 \ {\rm yr}$ or once $M_{pl} > 100 M_\oplus$. 
For a given $\beta$ and a given $a$ we then determine the value of $f_{dust}$ (which within 
our assumptions is a measure of the metallicity) for which a gas giant planet ($M_{pl} > 100 M_\oplus$) 
will form in exactly $5 \times 10^6$ yrs. This value of $f_{dust}$ ($f_{dust,min}$) is the minimum 
value for which a gas giant planet can form prior to the dissipation of the gas disk, and
defines a group of planets that should undergo very little Type II migration. 

\section{Results} 
For a protoplanetary disk with a specified value of $\beta$, we use the model 
described above to calculate the mimimum or threshold metallicity required to 
form a planet at radius $a$ prior to the dissipation of the gas disk. 
We consider radii between $1$ and $10$ AU and assume, initially, that the core 
and growing planet suffer negligible Type~I migration during the formation 
process. Illustrative runs for a planet growing at 5~AU in a disk with 
$\beta = 1.5$ are shown in Figure~\ref{planprofile}. As the metallicity is 
increased (i.e. larger values of $f_{\rm dust}$) growth of the planet 
toward the critical core mass is accelerated. In this case values of $f_{dust} = 2.05$ 
and $f_{dust} = 2.35$ fail to yield a fully-formed giant planet by our 
definition, as the planet mass after $5$ Myrs is less than $100 M_\oplus$. 
Slightly higher metallicity ($f_{\rm dust} = 2.45$) however results in 
runaway runaway growth and produces a $100 M_\oplus$ gas giant planet 
within 5 Myr.

By interpolating from a series of such runs in which the planet formation 
time scale brackets the assumed disk lifetime, we determine $f_{dust,min}$ 
as a function of orbital radius for each disk model. We consider surface 
density profiles of $\beta = 1$, $\beta = 1.5$, and $\beta = 2$, in each case 
normalizing the value of the planetesimal surface density at an 
arbitrary radius of 1~AU.

Figure \ref{nomig} shows the derived values of $f_{dust,min}$ against radius 
for the three planetesimal surface density profiles that we consider. The most 
obvious feature in the plots is the sharp rise in the predicted threshold 
metallicity interior to the radius of the snow line. Although the {\em absolute}  
values of the threshold metallicity obviously scale with the assumed 
normalization (and hence the values agree for all disk models at 1~AU), for 
all models there is a jump of around an order of magnitude across the 
snow line. Since observationally the frequency of planets around stars 
significantly more metal-poor than the Sun is very low \citep{gonzalez98,
santos01,santos04,fischer04,fischer05}, this implies that gaseous planets 
cannot form, {\em in situ}, within the snowline (here at 2.7~AU) unless
the metallicity is significantly higher than solar. This result is largely 
consistent with \citet{ida04}, who found that gaseous planets would only 
form within the snowline for large planetesimal surface densities. Although 
there are hints of a correlation between planet orbital period and host 
[Fe/H] \citep{sozzetti04}, it is clear that the observed distribution of 
extrasolar planets in the $a$-[Fe/H] plane does not resemble Figure~\ref{nomig}. 
There are a number of planets with semi-major axes $a < 1$~AU around metal-poor 
stars, and no clear sign of a jump at any plausible snow line radius. At 
relatively small orbital radii, then, migration appears to be necessary 
in order to explain the observed statistics of extrasolar planets.

Figure \ref{nomig} also shows the behavior of the threshold metallicity with 
orbital radius beyond the snow line. The minimum metallicity required for 
gas giant formation depends only weakly on radius out to $\sim 6$~AU for
all of the surface density profiles that we have considered, and for the 
flattest profile ($\beta = 1$) there is little dependence out to larger 
radii of around 10~AU. This is consistent with the model developed by 
\citet{pollack96}, which predicts that Jupiter forms at $5.2$ AU in $8$ Myr 
with a local planetesimal surface density of $10 \mathrm{ \ g \ cm^{-2}}$ 
while Saturn forms within a comparable time scale ($10$~Myr) if the 
local planetesimal surface density at 9.5~AU is $3 \mathrm{ \ g \ cm^{-2}}$. 
There is a weak trend for planet formation to be favored at smaller radii 
around lower metallicity stars \citep{pinotti05} if $\beta = 2$, but in 
general we would expect that there should not be a strong radial dependence 
over the range of orbital radii currently accessible to radial velocity 
surveys (i.e. the threshold metallicity for planet formation at $3$~AU 
should be similar to that at $6$~AU). At larger radii, however, there is 
a significant dependence which varies between models. The predicted threshold 
metallicity rises by a factor of $\approx 2$ between 6~AU and 10~AU if 
$\beta = 1$, whereas for a steeper $\beta = 2$ profile the rise is closer 
to a factor of 4. Within the context of core accretion models, measurement 
of the lower envelope of detected planets in the $a$-[Fe/H] plane at fairly 
large radii can therefore constrain some inputs to the formation model. 
Astrometric surveys appear to offer the best possibilities for 
assembling large enough planet samples at the desired radii. 
 
\subsection{Core migration}
The above calculations, like the baseline core accretion models presented by 
\citet{pollack96}, assume that the core grows at fixed orbital radius. If the 
core remains embedded within a gaseous disk, however, analytic calculations 
\citep{ward97} show that differential torques arising from the core's 
gravitational interaction with the gas should induce rapid Type~I migration. 
The influence of torques from turbulent fluctuations in the disk surface 
density may also drive orbital drift \citep{nelson04, laughlin04}. Migration 
in either regime can accelerate gaseous planet formation \citep{hourigan84,
alibert04,rice03}, but can also prevent planet formation if the cores 
migrate into the central star prior to the accretion of the envelope and 
subsequent gap opening. Here, we consider the possible influence of 
Type~I core migration on the expected distribution of massive planets in 
the orbital radius-host metallicity plot.

Preliminary calculations showed, as expected, that the survival prospects 
for cores allowed to migrate at the analytic Type~I rate \citep{ward97} are 
slim. To allow the core to survive, we therefore assume that the migration, 
on average, must be slower than the canonical Type I migration rate by about 
an order of magnitude. This assumption is similar to that made by \citet{alibert04}, 
and may be justified, in part, by numerical simulations that suggest that 
the true Type~I migration rate may be significantly slower than previously 
assumed \citep{miyoshi99,jangcondell05}. Specifically, we consider three 
different core migration rates,
\begin{eqnarray}
 \dot{a} &=& 4 \times 10^{-7} \ {\rm AU \ yr^{-1}} \\
 \dot{a} &=& 8 \times 10^{-7} \ {\rm AU \ yr^{-1}} \\
 \dot{a} &=& 1 \times 10^{-7} \left({M_{pl} \over M_\oplus}\right) 
 \left({a \over \mathrm{AU}}\right)^{1/2} \ {\rm AU \ yr^{-1}}.
\label{eq_rates} 
\end{eqnarray}   
These prescriptions are intended only to sample a range of migratory behavior 
that allows giant planets to form without being consumed as cores by the star. 
We do not suggest that any of these rates is representative of the actual core 
migration rate, although the latter rate has a mass dependence
that matches the standard Type~I migration rate. Since we adjust the 
planetesimal surface density in the feeding zone as the core grows, it is 
straightforward to add core migration to the model described in section \S2. 
As before, we vary $f_{dust}$ to determine the threshold metallicity required 
to form a planet at {\em final} radius $a_{final}$ in $5 \times 10^6$ yr. This 
requires only an additional iteration to determine the (initially unknown) 
value of $a_{initial}$ that yields such a planet. Henceforth we consider 
only surface density profiles of $\beta = 1.5$. 

Figure \ref{mig} shows $f_{dust,min}$ against $a_{final}$ for the three migration rates 
that we considered (\ref{eq_rates}). The results for all three migration prescriptions 
are qualitatively similar. The prominent jump in threshold metallicity at the snow line, 
seen in the calculations with a static core, is erased in the case of a migrating core 
which can accrete planetesimals across a much wider range of orbital radii. These 
smaller threshold values interior to the snow line appear to be in better accord 
with observations. Moreover, at larger radii there is now a steady increase in the 
predicted threshold metallicity with radius, even at radii accessible to ongoing 
radial velocity surveys for extrasolar planets. If Type~I migration is implicated 
in the formation of gas giants, we would therefore expect to see a steady rise 
in the minimum host metallicity as the orbital radius of the planet increases.

Figure \ref{afin} shows $a_{initial}$ against $a_{final}$ for the non-constant migration rate,
and illustrates the amount of migration that has taken place. 
For planets that remain beyond the snowline ($a_{final} > 2.7$ AU), the
change in semi-major axis is largely independent of radius ($\Delta a \sim 3$ AU). The radial
dependence of the migration rate appears to be balanced by the radial dependence of planet's 
growth rate. For planets that end up within the snowline ($a_{final} < 2.7$ AU), $a_{initial}$
appears to be almost constant, with $\Delta a \sim 4$ AU for $a_{final} = 1$ AU, and $\Delta
a \sim 3$ AU for $a_{final} = 2$ AU. This suggests that the surface density discontinuity at the
snowline can produce a large change in $a_{final}$ for a small change in $a_{initial}$. Although
our chosen migration rates are somewhat arbitrary, these results suggest that migration rates of
between $2$ and $4$ AU in $5 \times 10^6$ yrs can produce planets within the snowline around stars with
reasonably low metallicities, and can result in an increase in the threshold metallicity with
radius.  

By focusing on the shape of the critical metallicity curve, we have deliberately 
avoided detailed discussion of the {\em relative number} of planets expected to populate 
different regions of the [Fe/H]-$a$-$M_{pl}$ space. This depends upon the fraction 
of plausible initial conditions that yield each specific outcome \citep{armitage02,ida04}. 
We note, however, that such statistical considerations may yield additional 
evidence for Type~I migration of gas giant cores. In particular, \citet{ida04} predict 
a dearth of planets with masses between $10$ and $100 M_\oplus$ within the snowline, 
a region they term the ``planet desert''. In their model, a planet can only fall 
within the desert if its growth is halted during the rapid envelope growth phase. 
This is an unlikely outcome since this phase lasts for such a short time. Equivalently, 
in the absence of core migration the range of metallicity values $\Delta f_{dust}$ 
that populate the desert region is small compared to $f_{dust,min}$, and as a 
consequence few planets are predicted in the desert.

When core migration is included, this prediction of an unoccupied planet desert can 
be modified, though the results do depend on the details of the migration history. As 
shown in Figure \ref{mig}, $f_{dust,min}$ at small orbital radii is greatly reduced 
in the presence of migration. Compared to this threshold value, we find a reasonably 
large range of $f_{dust}$ values that result in planet masses that fall within the 
``planet desert''. For the non-constant migration rate, 
$f_{dust} = 1.3$ yields a $100 M_\oplus$ planet at $1$ AU within $5$ Myr, while
$f_{dust} = 1.1$ produces a $48 M_\oplus$ planet after $5$ Myr. Roughly 20\% of 
stars with metallicities close to the threshold value would then be candidates for 
forming planets in the otherwise (in the absence of migration) unoccupied region 
of phase space. Unfortunately, the range of $f_{dust}$ values that resulted in
planets with masses within the desert region was smaller for the constant migration 
rates than for the non-constant migration rate. This suggests that for planets to lie 
within the desert, the core migration rate must be such as to maintain a large 
planetesimal accretion rate, $\dot{M}_{\rm core}$, resulting in a large critical core 
mass (see equation \ref{Mcrit}). Therefore, although the existence of planets 
within the desert region may point to the role of Type~I migration, more 
detailed modeling will be required to establish this clearly. 

\section{Application to current data}
To compare these expectations with current observations of extrasolar 
planetary systems, we use the metallicity data published by \cite{santos04}. 
The \cite{santos04} catalog includes [Fe/H] measurements, derived using 
uniform analysis methods, for 98 planet host stars. For our purposes we 
include all planets in multiple planet systems, and take the average value 
of [Fe/H] for those stars with more than one derived abundance. HD47536b and 
its host star are excluded due to the large uncertainty in the mass of that 
planet, leaving 110 massive planets for which we have the host star 
metallicity, together with the planet mass $m_p \sin(i)$, semi-major axis and 
eccentricity. Fig.~\ref{FeHvsa} shows the distribution of these planets 
in the $a$-[Fe/H] plane. Although there are some hints that this distribution 
is not merely a scatter plot (for example there are no observed extrasolar 
planets at $a > 1 \ {\rm AU}$ around stars with [Fe/H]~$<$-0.4), the lower 
envelope of the distribution which has been the focus of the theoretical 
discussion is obviously poorly defined in current data due to the small 
number of planets observed that are unambiguously outside the snow line. 
Nevertheless, we can use the data for the more limited purpose of testing 
whether some of our basic assumptions are consistent with observations.

Theoretically, we expect that planets observed near the critical metallicity 
line formed late in the 
lifetime of the gaseous protoplanetary disk. We would expect them to have 
lower masses -- since there was less time available to accrete gas before 
the disk was dissipated -- and possibly different eccentricity. To test 
whether these expectations are consistent with current data, we divide 
the planets with semi-major axis $a > 0.5$~AU (i.e. excluding the hot 
Jupiters, but probably including some planets that are now seen interior 
to the snow line) into high and low host star metallicity samples. Some 
care is necessary, because the existing sample of planets discovered via 
radial velocity is biased against the discovery of low mass planets at 
large radii. Specifically, for a fixed number of radial velocity 
measurements with a given noise level (and some implicit assumptions as
to the time sampling of the survey), the minimum detectable planet 
mass will scale with orbital radius roughly as $[m_p \sin(i)]_{\rm min} 
\propto a^{1/2}$. This detection bias means that simply dividing 
the sample of planets into two -- those with host stars above and below some 
fixed value of [Fe/H] -- risks mixing a variation of host metallicity with 
planet orbital radius into a spurious trend of planet mass with host 
metallicity.

To avoid such potential biases, we construct matched samples of planets 
around high and low metallicity host stars. We first bin planets 
according to their semi-major axis (in 0.5~AU increments between 
0.5~AU and 3~AU, plus one final bin from 3-4~AU). Within each radial bin, 
the minimum detectable planet mass in a radial velocity survey 
varies by much less than the intrinsic dispersion in observed planet masses. 
We then split the planets {\em within each radial bin} into 
subsamples of high and low metallicity host stars, with (as far as is 
possible) equal numbers of planets in each. Our final sample of planets 
around relatively low (high) metallicity stars then consists of all the 
planets below (above) the dashed lines shown in Fig.~\ref{FeHvsa}.  

The mass distribution of the planets in the high and low metallicity 
samples, defined as above, is shown in Fig.~\ref{fig_KS}. The two 
distributions are consistent with the expectation that the planets 
around lower metallicity stars have, on average, lower masses. Formally, 
a Kolmogorov-Smirnov test shows that the probability that the two 
distributions are drawn from the same parent distribution is 
$P_{\rm KS} = 6 \times 10^{-3}$, so the statistical evidence in support 
of the existence of the hypothesized `no-migration' curve in the $a$-[Fe/H] 
plane is currently suggestive rather than overwhelming. However, the fact 
that some indications of a signal are present in current data does suggest 
that the larger planet samples that will plausibly be accumulated in the 
near future should suffice to test some of the ideas outlined in the 
preceding Section.

Mindful of the fact that in some theoretical models orbital migration 
is associated with concurrent eccentricity growth \citep{papaloizou01,
murray02,goldreich03,ogilvie03}, we have tested whether the eccentricity 
distribution of the high and low metallicity samples is significantly 
different. Unlike in the case of the $m_p \sin(i)$ distributions, the 
eccentricity distributions of the two samples are statistically 
indistinguishable. Within the context of our model there 
is no evidence that relatively early planet formation, followed by 
significant radial migration, promotes growth of the final eccentricity.

\section{Summary}
In this paper we have investigated how, with the addition of knowledge 
of the host stars' metallicity, the statistics of extrasolar planets 
can be used to constrain models for giant planet formation. Our main 
results are:
\begin{itemize}
\item[1] 
By studying planets at a particular radius around the lowest metallicity 
host stars, it is possible to isolate a subsample that will have suffered, 
on average, less Type~II migration than the typical planet at that 
radius. This separation will only be clean outside the snow line, and 
only if metallicity is the most important random variable affecting the 
time scale for core formation. Comparison of such a non-migratory 
sample with planets around metal-rich stars could constrain the amount 
of mass accreted onto the planet during Type~II migration.
\item[2]
If giant planet cores grow in place, it is difficult to explain the 
presence of massive extrasolar planets around relatively low metallicity  
hosts within the snow line. Static core growth models predict a threshold 
metallicity for planet formation that is roughly flat within 6~AU, but which 
rises to larger radius. The functional dependence varies according to the slope 
of the planetesimal surface density distribution.
\item[3]
If giant planet cores suffer Type~I migration as they grow, the 
threshold metallicity rises smoothly beyond $a \approx 2$~AU. The 
details of the migration are relatively unimportant, provided that 
the overall rate is slow compared to the canonical analytic 
predictions. Under limited conditions, it is possible to populate 
what would otherwise be a desert in the distribution of planets 
with masses $10 \ M_\oplus < M_{pl} < 100 \ M_\oplus$ at small 
orbital radii \citep{ida04}.
\end{itemize}
Existing observations appear to be consistent with the basic theoretical 
premise of this paper---that host metallicity is the dominant factor 
controlling the time scale for massive planet formation. This hypothesis 
is motivated by the observed dependence of planet frequency on stellar 
[Fe/H] \citep{gonzalez98,santos01,santos04,fischer04,fischer05}, and 
implies a planet mass - metallicity correlation that is seen, albeit 
at low significance, in the data. This leaves us hopeful that with 
larger samples of massive extrasolar planets useful constraints on 
planet formation models will be attainable. 

\acknowledgements

This work was supported by NASA under grants NAG5-13207 and NNG04GL01G 
from the Origins of Solar Systems and Astrophysics Theory Programs, and 
by the NSF under grant AST~0407040. The hospitality of the Aspen Center for 
Physics, where part of this paper was completed, is gratefully acknowledged.

\newpage

\begin{figure}
\plotone{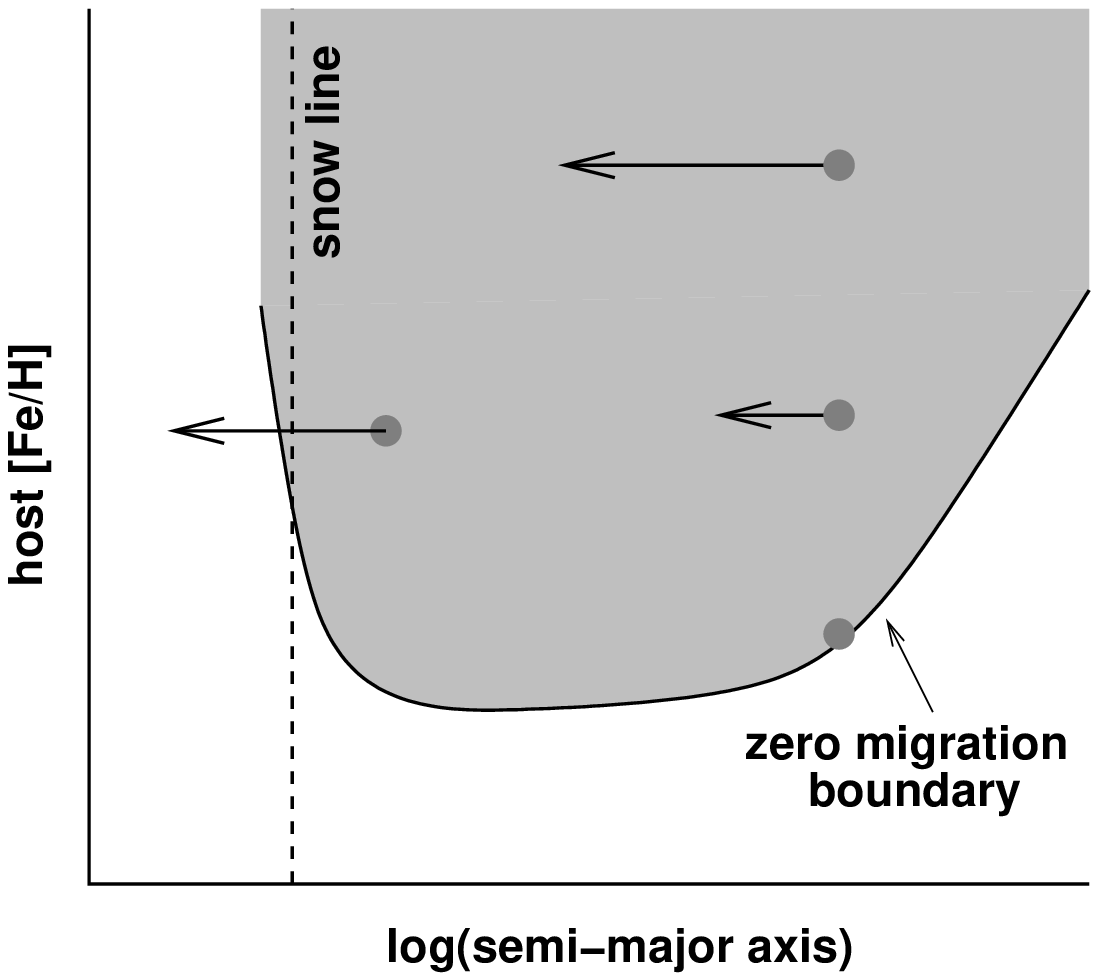}
\figcaption{Illustration of the expected influence of core formation 
and orbital migration on the final distribution of extrasolar planets. 
In an idealized model, in which the initial mass and lifetime of the gas 
disk have a narrow range of values, giant planet formation will occur at 
a given radius provided that the host star's metallicity exceeds a threshold
value. At this threshold value, runaway occurs and the gaseous envelope is 
accreted just as the protoplanetary disk is about to be dissipated, allowing 
no opportunity for Type~II migration. For higher host metallicities, Type~II 
migration is expected to become increasingly important. If the threshold 
metallicity is an increasing function of radius -- in practice at radii 
outside the snow line -- then inward migration {\em cannot} populate the 
unshaded region below the critical curve. The shape of this curve then defines 
a threshold for giant planet formation that is independent of the effects 
of orbital migration.}
\label{fig_schematic}
\end{figure}

\begin{figure}
\plotone{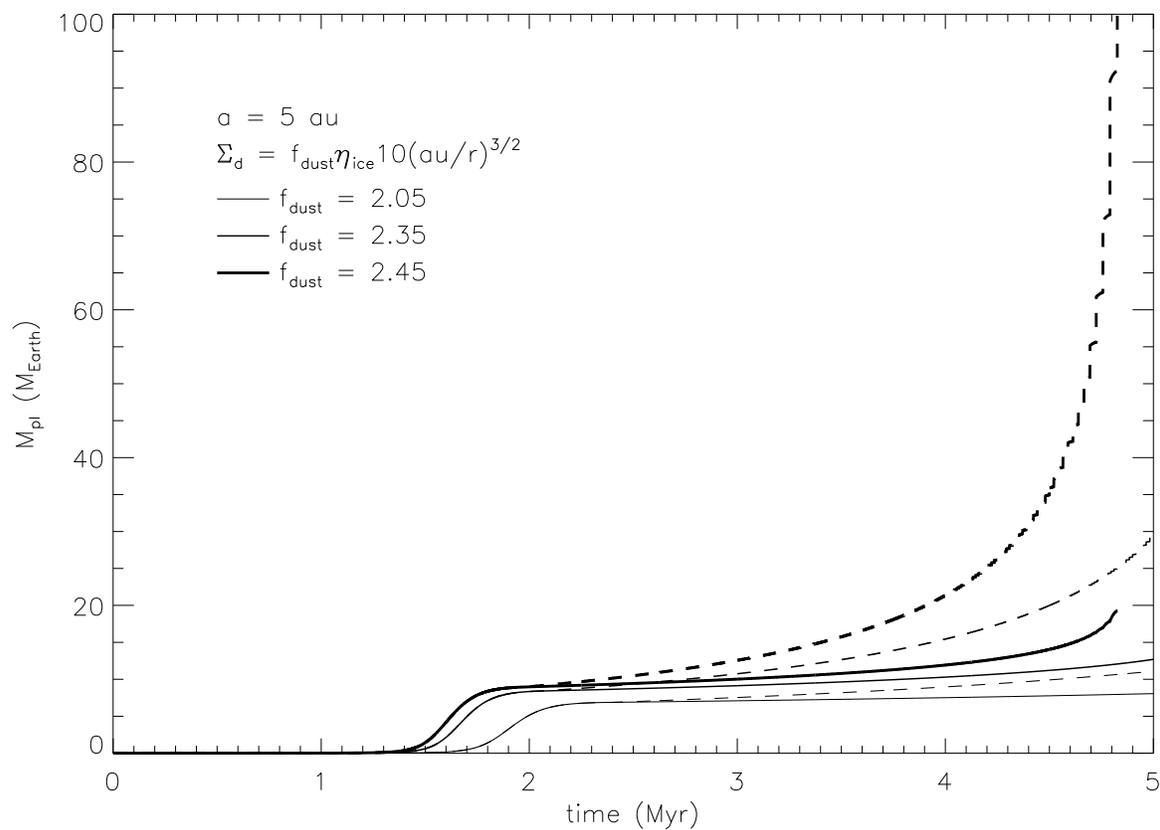}
\caption{Growth curves for a core located at $a = 5$ AU and for three
different planetesimal surface densities, as defined by $f_{dust}$. The different
$f_{dust}$ values are represented by different line thicknesses, and in each case
the solid line shows the core mass while the dashed line is the total planet mass (core + envelope).
For $f_{dust} = 2.45$
a giant planet forms within $5$ Myr, while for lower values of $f_{dust}$ gaseous planet
formation does not occur within the disk lifetime.}
\label{planprofile}
\end{figure}

\begin{figure}
\plotone{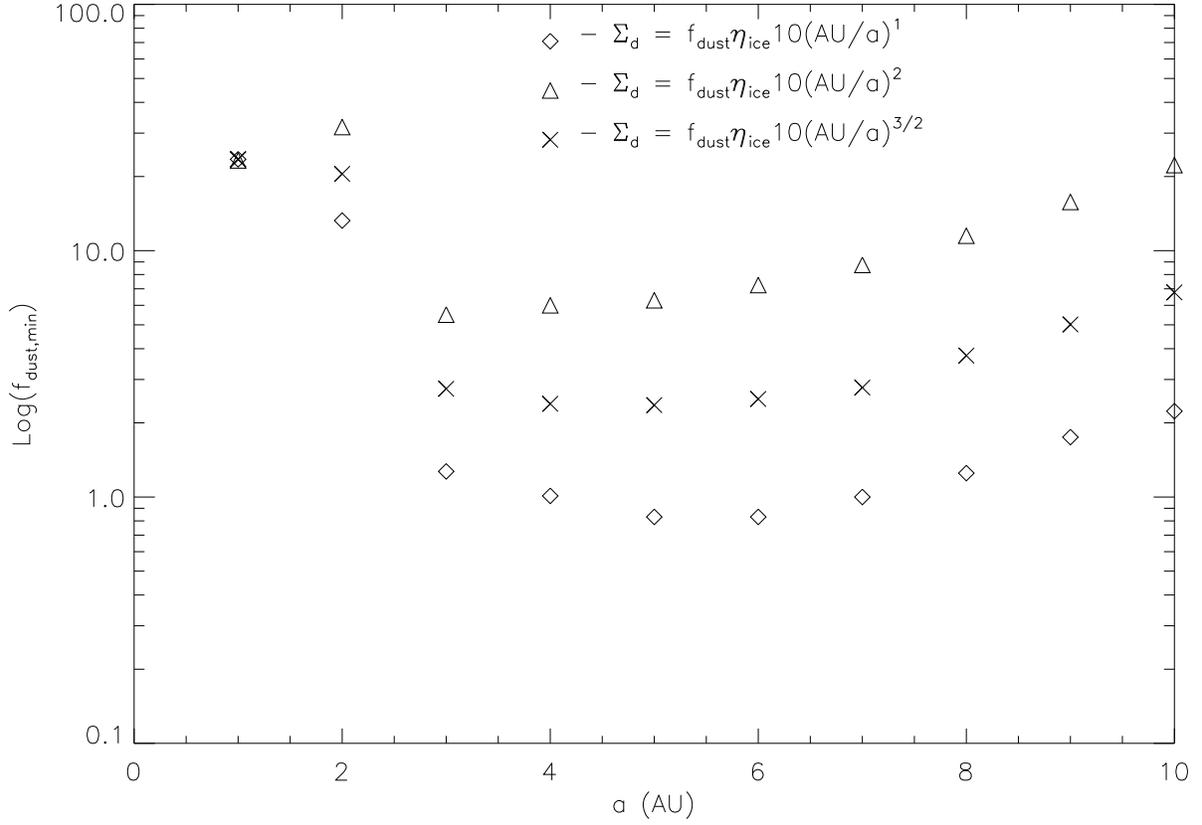}
\caption{The minimum value of $f_{dust}$ required to form a gaseous planet, {\em in situ}, within $5$ Myr.
This illustrates the difficulty in forming gaseous planets within the snowline ($\sim 2.7$ AU), since the
required planetesimal surface density is extremely high, and shows that even for reasonably steep surface density profiles,
the radial dependence beyond the snowline is relatively weak out to $\sim 7$ AU. }
\label{nomig}
\end{figure}

\begin{figure}
\plotone{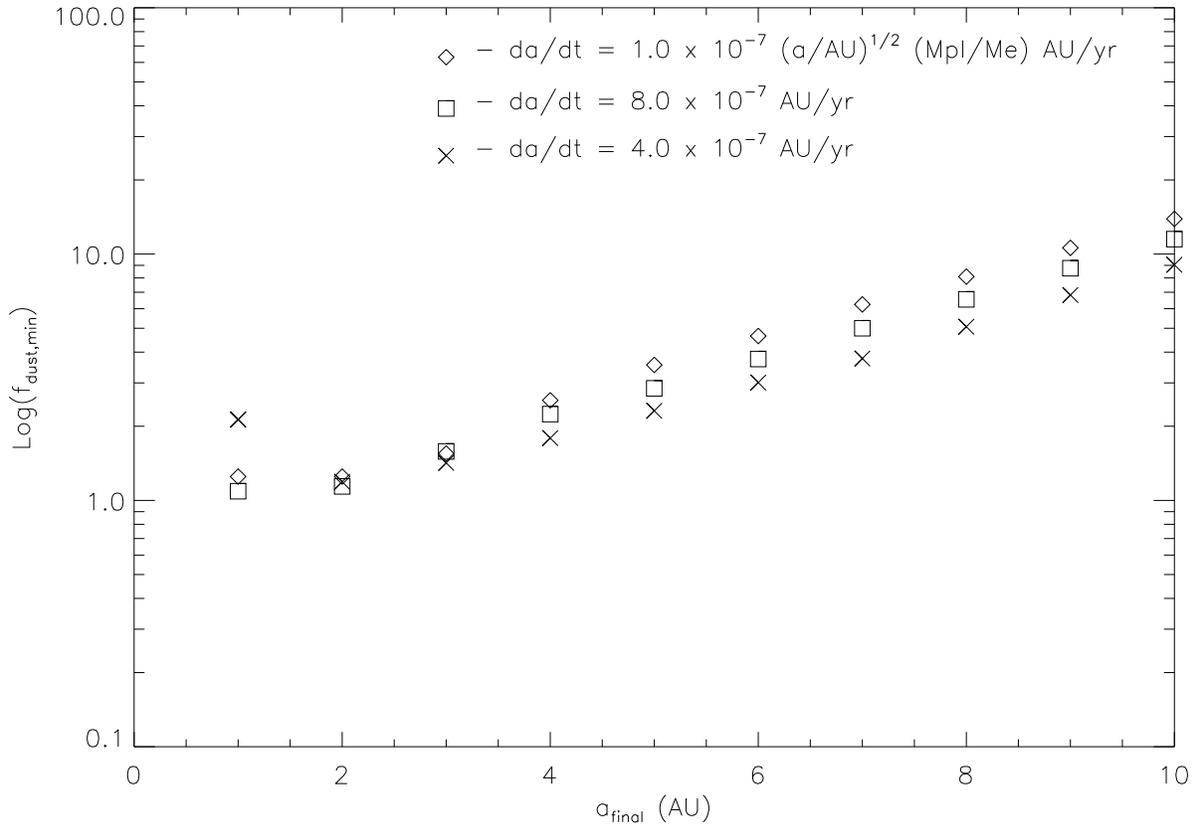}
\caption{The minimum value of $f_{dust}$ required to form a gaseous planet if the core is assumed to
migrate to a final radius $a_{final}$. We consider three different migration rates and find that
in all cases gaseous planets can have final semi-major axes within the snowline for $f_{dust}$ values that
are significantly smaller than that required when core migration is ignored.  This figure also suggests that
if core migration plays a role in gaseous planet formation the metallicity required may increase with increasing
$a_{final}$.}
\label{mig}
\end{figure}

\begin{figure}
\plotone{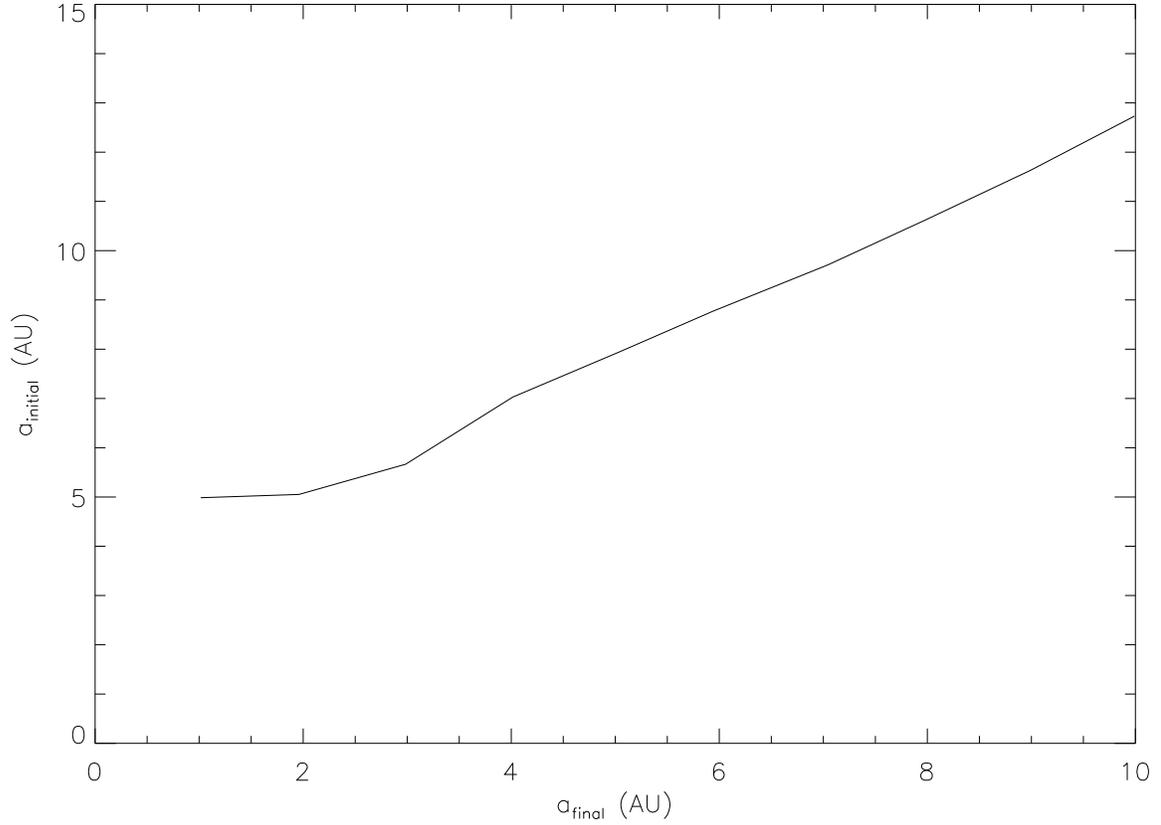}
\caption{Initial semi-major axis ($a_{initial}$) against final semi-major axis ($a_{final}$) for
the non-constant migration rate. For planets that remain beyond the snowline the change in semi-major axis
is almost constant despite the radial dependence of the migration rate.  For planets that end up
within the snowline, $a_{initial}$ appears almost constant suggesting that the surface density
discontinuity at the snowline can produce large changes in $a_{final}$ for small changes in $a_{initial}$.}
\label{afin}
\end{figure}

\begin{figure}
\plotone{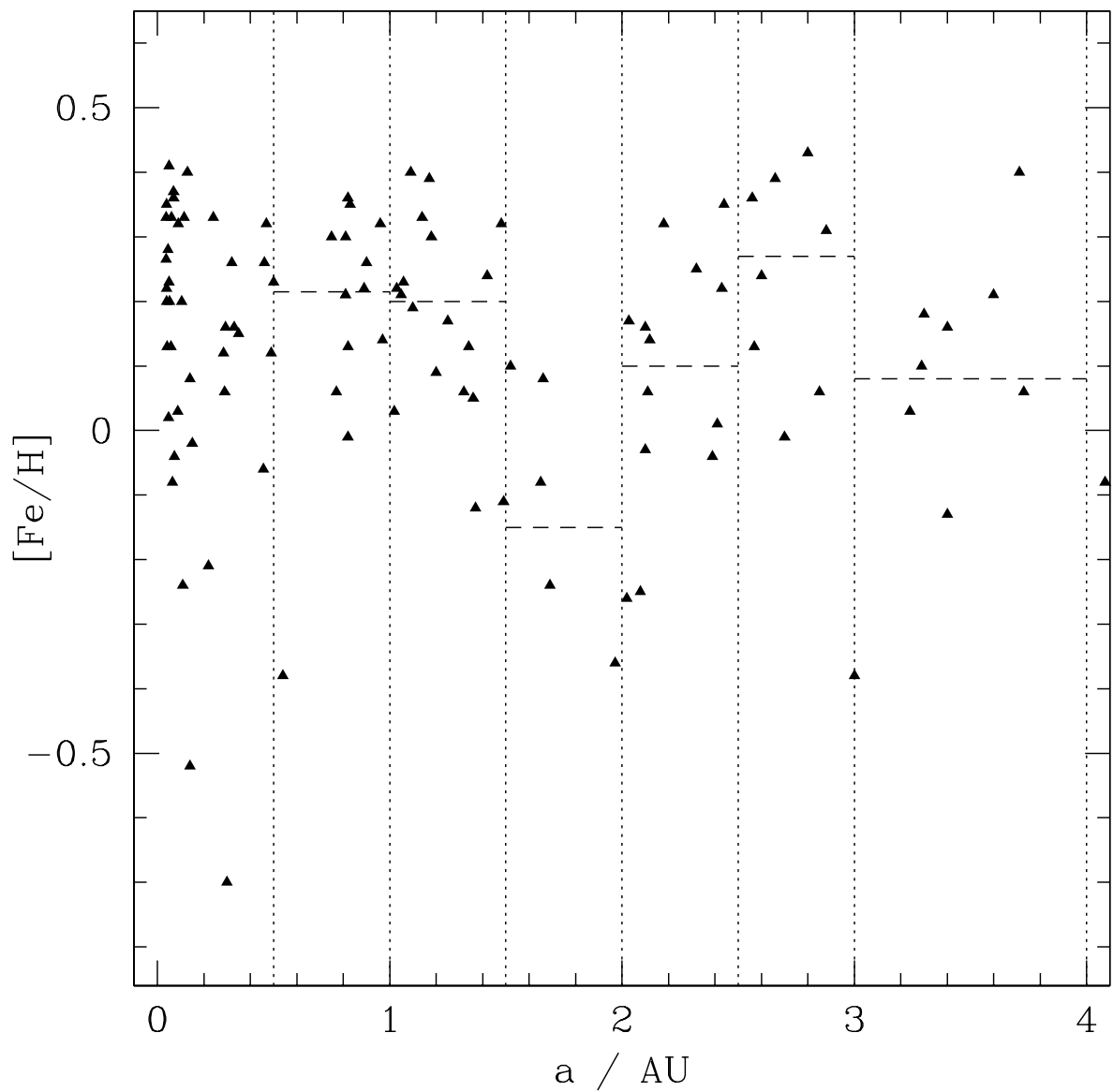}
\caption{Distribution of extrasolar planets in $a$-[Fe/H] plane. Although there
is insufficient data to strongly constrain the relationship between semi-major axis and
metallicity, there is a hint that the metallicity required for giant planet formation
increases with increasing radius. The divisions within the figure are used to divide the
planets within each radius bin into high- and low-metallicity samples.} 
\label{FeHvsa}
\end{figure}

\begin{figure}
\plotone{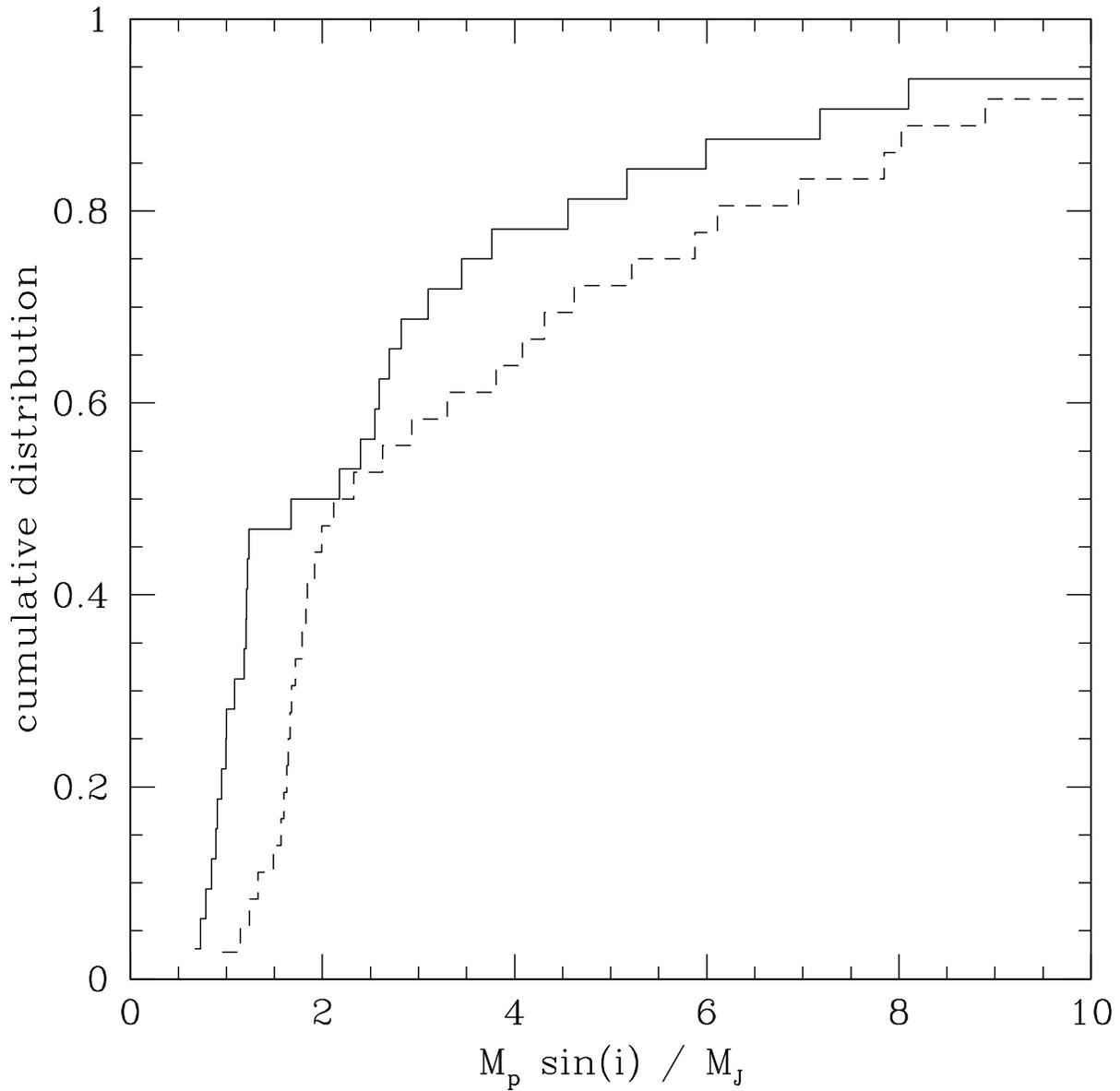}
\caption{Mass distribution of the planets in the high and low metallicity samples defined by
the divisions in Figure \ref{FeHvsa}. The two distributions are consistent with the expectation that
the planets around lower metallicity stars (dashed line) have, on average, lower masses. This is
consistent with the idea that planets around lower metallicty stars are likely to have formation times 
comparable to the disk lifetimes, and are unlikely to undergo significant Type II migration.}
\label{fig_KS}
\end{figure}

\end{document}